# The Contestation of Tech Ethics: A Sociotechnical Approach to Technology Ethics in Practice

Ben Green*

**Abstract:** This article introduces the special issue "Technology Ethics in Action: Critical and Interdisciplinary Perspectives". In response to recent controversies about the harms of digital technology, discourses and practices of "tech ethics" have proliferated across the tech industry, academia, civil society, and government. Yet despite the seeming promise of ethics, tech ethics in practice suffers from several significant limitations: tech ethics is vague and toothless, has a myopic focus on individual engineers and technology design, and is subsumed into corporate logics and incentives. These limitations suggest that tech ethics enables corporate "ethics-washing": embracing the language of ethics to defuse criticism and resist government regulation, without committing to ethical behavior. Given these dynamics, I describe tech ethics as a terrain of contestation where the central debate is not whether ethics is desirable, but what "ethics" entails and who gets to define it. Current approaches to tech ethics are poised to enable technologists and technology companies to label themselves as "ethical" without substantively altering their practices. Thus, those striving for structural improvements in digital technologies must be mindful of the gap between ethics as a mode of normative inquiry and ethics as a practical endeavor. In order to better evaluate the opportunities and limits of tech ethics, I propose a sociotechnical approach that analyzes tech ethics in light of who defines it and what impacts it generates in practice.

## 1 Introduction: A Crisis of Conscience

If digital technology production in the beginning of the 2010s was characterized by the brash spirit of Facebook's motto "move fast and break things" and the superficial assurances of Google's motto "do not be evil", digital technology toward the end of the decade was characterized by a "crisis of conscience"[1]. While many have long been aware of digital technology's harms, an influx of stories about salient harms led to widespread critique of digital technology. The response was the "techlash": a growing public animosity toward major technology companies. In 2018, Oxford Dictionaries and the Financial Times both deemed techlash to be one of the words of the year[2, 3].

Consider just a few of the controversies that prompted this crisis of conscience within tech and the associated techlash:

**Disinformation:** Throughout the 2016 US presidential election between Donald Trump and Hillary Clinton, social media was plagued with fraudulent stories that went viral[4, 5]. In turn, numerous commentators—including Hillary Clinton—blamed Facebook for Donald Trump's presidential election victory[6–9]. Later reporting revealed that Facebook's leadership has actively resisted taking strong measures to curb disinformation, instead prioritizing the company's business strategies[10, 11].

**Cambridge Analytica:** In 2018, *The New York Times*

● Ben Green is with the Society of Fellows and the Gerald R. Ford School of Public Policy, University of Michigan, Ann Arbor, MI 48109, USA. E-mail: bzgreen@umich.edu.
∗ To whom correspondence should be addressed.
   Manuscript received: 2021-05-20; accepted: 2021-10-20

and *The Guardian* reported that the voter-profiling firm Cambridge Analytica had harvested information from millions of Facebook users, without their knowledge or permission, in order to target political ads for Donald Trump's 2016 presidential campaign[12, 13]. Cambridge Analytica had acquired these data by exploiting the sieve-like nature of Facebook's privacy policy.

**Military and ICE Contracts**: In 2018, journalists revealed that Google was working with the US Department of Defense (DoD) to develop software that analyzes drone footage[14]. This effort, known as Project Maven, was part of a ＄7.4 billion investment in AI by the DoD in 2017[14] and represented an opportunity for Google to gain billions of dollars in future defense contracts[15]. Another story revealed that Palantir was developing software for Immigration and Customs Enforcement (ICE) to facilitate deportations[16].

**Algorithmic Bias:** In 2016, ProPublica revealed that an algorithm used in criminal courts was biased against Black defendants, mislabeling them as future criminals at twice the rates of white defendants[17]. Through popular books about the harms and biases of algorithms in settings such as child welfare, online search, and hiring[18−20], the public began to recognize algorithms as both fallible and discriminatory.

These and other tech-related controversies were a shock to many, as they arrived in an era of widespread (elite) optimism about the beneficence of technology. Yet these controversies also brought public attention to what scholars in fields such as Science, Technology, and Society (STS), philosophy of science, critical data and algorithm studies, and law have long argued: technology is shaped by social forces, technology structures society often in deleterious ways, and technology cannot solve every social problem. Broadly speaking, these fields bring a "sociotechnical" approach to studying technologies, analyzing how technologies shape, are shaped by, and interact with society[21−24]. As tech scandals mounted, a variety of sociotechnical insights, long ignored by most technologists and journalists, were newly recognized (or in some form recreated).

Many in the tech sector and academia saw the harms of digital technology as the result of an inattention to ethics. On this view, unethical technologies result from a lack of training in ethical reasoning for engineers and a dearth of ethical principles in engineering practice[1, 25−28]. In response, academics, technologists, companies, governments, and more have embraced a broad set of goals often characterized with the label "tech ethics": the introduction of ethics into digital technology education, research, development, use, and governance. In the span of just a few years, tech ethics has become a dominant discourse discussed in technology companies, academia, civil society organizations, and governments.

This article reviews the growth of tech ethics and the debates that this growth has prompted. I first describe the primary forms of tech ethics in practice. I focus on the people and organizations that explicitly embrace the label of "tech ethics" (and closely related labels, such as AI ethics and algorithmic fairness). I then summarize the central critiques made against these efforts, which call into question the effects and desirability of tech ethics. Against the backdrop of these critiques, I argue that tech ethics is a terrain of contestation: the central debate is not whether ethics is desirable but what ethics entails and who has the authority to define it. These debates suggest the need for a sociotechnical approach to tech ethics that focuses on the social construction and real-world effects of tech ethics, disambiguating between the value of ethics as a discipline and the limits of tech ethics as a practical endeavor. I introduce this approach through four frames: objectivity and neutrality, determinism, solutionism, and sociotechnical systems.

## 2　The Rise of Tech Ethics

Although some scholars, activists, and others have long considered the ethics of technology, attention to digital technology ethics has rapidly grown across the tech industry, academia, civil society, and government in recent years. As we will see, tech ethics typically involves applied forms of ethics such as codes of ethics and research ethics, rather than philosophical inquiry (i.e., moral philosophy). For instance, one common treatment of tech ethics is statements of ethical principles. One analysis of 36 prominent AI principles documents shows the sharp rise in these statements, from 2 in 2014 to 16 in 2018[29]. These documents tend to cover the themes of fairness and non-discrimination, privacy, accountability, and transparency and explainability[29]. Many documents also reference human rights, with some taking international human rights as the framework for ethics[29].

## 2.1 Tech industry

The most pervasive treatment of tech ethics within tech companies has come in the form of ethics principles and ethics oversight bodies. Companies like Microsoft, Google, and IBM have developed and publicly shared AI ethics principles, which include statements such as "AI systems should treat all people fairly" and "AI should be socially beneficial"[30−32]. These principles are often supported through dedicated ethics teams and advisory boards within companies, with such bodies in place at companies including Microsoft, Google, Facebook, DeepMind, and Axon[33−37]. Companies such as Google and Accenture have also begun offering tech ethics consulting services[38, 39].

As part of these efforts, the tech industry has formed several coalitions aimed at promoting safe and ethical artificial intelligence. In 2015, Elon Musk and Sam Altman created OpenAI, a research organization that aims to mitigate the "existential threat" presented by AI, with more than ＄1 billion in donations from major tech executives and companies[40]. A year later, Amazon, Facebook, DeepMind, IBM, and Microsoft founded the Partnership on AI (PAI), a nonprofit coalition to shape best practices in AI development, advance public understanding of AI, and support socially beneficial applications of AI[41, 42].①

## 2.2 Academia

Computer and information science programs at universities have rapidly increased their emphasis on ethics training. While some universities have taught computing ethics courses for many years[44−46], the emphasis on ethics within computing education has increased dramatically in recent years[47]. One crowdsourced list of tech ethics classes contains more than 300 courses[48]. This plethora of courses represents a dramatic shift in computer science training and culture, with ethics becoming a popular topic of discussion and study after being largely ignored by the mainstream of the field just a few years prior.

Research in computer science and related fields has also become more focused on the ethics and social impacts of computing. This trend is observable in the recent increase in conferences and workshops related to computing ethics. The ACM Conference on Fairness, Accountability, and Transparency (FAccT) and the AAAI/ACM Conference on AI, Ethics, and Society (AIES) both held their first annual meetings in February 2018 and have since grown rapidly. There have been several dozen workshops related to fairness and ethics at major computer science conferences[49]. Many universities have supported these efforts by creating institutes focused on the social implications of technology. 2017 alone saw the launch of the AI Now Institute at NYU[50], the Princeton Dialogues on AI and Ethics[51], and the MIT/Harvard Ethics and Governance of Artificial Intelligence Initiative[52]. More recently formed centers include the MIT College of Computing[53]; the Stanford Institute for Human-Centered Artificial Intelligence[54]; and the University of Michigan Center of Ethics, Society, and Computing[55].

## 2.3 Civil society

Numerous civil society organizations have coalesced around tech ethics, with strategies that include grantmaking and developing principles. Organizations such as the MacArthur and Ford Foundations have begun exploring and making grants in tech ethics[56]. For instance, the Omidyar Network, Mozilla Foundation, Schmidt Futures, and Craig Newmark Philanthropies partnered on the Responsible Computer Science Challenge, which awarded ＄3.5 million between 2018 and 2020 to support efforts to embed ethics into undergraduate computer science education[57]. Many foundations also contribute to the research, conferences, and institutes that have emerged in recent years.

Other organizations have been created or have expanded their scope to consider the implications and governance of digital technologies. For example, the American Civil Liberties Union (ACLU) has begun hiring technologists and is increasingly engaged in debates and legislation related to new technology. Organizations such as Data & Society, Upturn, the Center for Humane Technology, and Tactical Tech study the social implications of technology and advocate for improved technology governance and design practices.

Many in civil society call for engineers to follow an ethical oath modeled after the Hippocratic Oath (an ethical oath taken by physicians)[20, 58−60]. In 2018, for instance, the organization Data for Democracy partnered

---

① Although PAI also includes civil society partners, these organizations do not appear to have significant influence. In 2020, the human rights organization Access Now resigned from PAI, explaining that "there is an increasingly smaller role for civil society to play within PAI" and that "we did not find that PAI influenced or changed the attitude of member companies"[43].

with Bloomberg and the data platform provider BrightHive to develop a code of ethics for data scientists, developing 20 principles that include "I will respect human dignity" and "It is my responsibility to increase social benefit while minimizing harm"[61]. Former US Chief Data Scientist DJ Patil described the event as the "Constitutional Convention" for data science[58]. A related effort, produced by the Institute for the Future and the Omidyar Network, is the Ethical OS Toolkit, a set of prompts and checklists to help technology developers "anticipate the future impact of today's technology" and "not regret the things you will build"[62].

## 2.4 Government

Many governments developed commissions and principles dedicated to tech ethics. In the United States, for example, the National Science Foundation formed a Council for Big Data, Ethics, and Society[63]; the National Science and Technology Council published a report about AI that emphasized ethics[64]; and the Department of Defense adopted ethical principles for AI[65]. Elsewhere, governing bodies in Dubai[66], Europe[67], Japan[68], and Mexico[69], as well as international organizations such as the OECD[70], have all stated principles for ethical AI.

## 3 The Limits of Tech Ethics

Alongside its rapid growth, tech ethics has been critiqued along several lines. First, the principles espoused by tech ethics statements are too abstract and toothless to reliably spur ethical behavior in practice. Second, by emphasizing the design decisions of individual engineers, tech ethics overlooks the structural forces that shape technology's harmful social impacts. Third, as ethics is incorporated into tech companies, ethical ideals are subsumed into corporate logics and incentives. Collectively, these issues suggest that tech ethics represents a strategy of technology companies "ethics-washing" their behavior with a façade of ethics while largely continuing with business-as-usual.

## 3.1 Tech ethics principles are abstract and toothless

Tech ethics codes deal in broad principles[71]. In 2016, for example, Accenture published a report explicitly outlining "a universal code of data ethics"[72]. A 2019 analysis of global AI ethics guidelines found 84 such documents, espousing a common set of broad principles: transparency, justice and fairness, non-maleficence, responsibility, and privacy[73]. Professional computing societies also present ethical commitments in a highly abstract form, encouraging computing professionals "to be ever aware of the social, economic, cultural, and political impacts of their actions" and to "contribute to society and human well-being"[74]. Ethics codes in computing and information science are notably lacking in explicit commitments to normative principles[74].

The emphasis on universal principles papers over the fault lines of debate and disagreement spurred the emergence of tech ethics in the first place. Tech ethics principles embody a remarkable level of agreement: two 2019 reports on global AI ethics guidelines noted a "global convergence"[73] and a "consensus"[29] in the principles espoused. Although these documents tend to reflect a common set of global principles, the actual interpretation and implementation of these principles raise substantive conflicts[73]. Furthermore, these principles have been primarily developed in the US and UK, with none from Africa or South America[73]. The superficial consensus around abstract ideals may thus hinder substantive deliberation regarding whether the chosen values are appropriate, how those values should be balanced in different contexts, and what those values actually entail in practice.

The abstraction of tech ethics is particularly troubling due to a lack of mechanisms to enact or enforce the espoused principles. When framed at such a high level of abstraction, values such as fairness and respect are unable to guide specific actions[75]. In companies, ethics oversight boards and ethics principles lack the authority to veto projects or require certain behaviors[76, 77]. Similarly, professional computing organizations such as the IEEE and ACM lack the power to meaningfully sanction individuals who violate their codes of ethics[75]. Moreover, unlike fields such as medicine, which has a strong and established emphasis on professional ethics, computing lacks a common aim or fiduciary duty to unify disparate actors around shared ethical practices[75]. All told, "Principles alone cannot guarantee ethical AI"[75].

## 3.2 Tech ethics has a myopic focus on individual engineers and technology design

Tech ethics typically emphasizes the roles and responsibilities of engineers, paying relatively little attention to the broader environments in which these individuals work. Although professional codes in computing and related fields assert general commitments to the public, profession, and one's employer, "the morality of a profession's or an employer's motives are not scrutinized"[74]. Similarly, ethics within computer science curricula tends to focus on ethical decision making for individual engineers[78].

From this individualistic frame comes an emphasis on appealing to the good intentions of engineers, with the assumption that better design practices and procedures will lead to better technology. Ethics becomes a matter of individual engineers and managers "doing the right thing" "for the right reasons"[79]. Efforts to provide ethical guidance for tech CEOs rest on a similar logic: "if a handful of people have this much power—if they can, simply by making more ethical decisions, cause billions of users to be less addicted and isolated and confused and miserable—then, is not that worth a shot?"[1]. The broader public beyond technical experts is not seen as having a role in defining ethical concerns or shaping the responses to these concerns[71].

Tech ethics therefore centers debates about how to build better technology rather than whether or in what form to build technology (let alone who gets to make such decisions). Tech ethics follows the assumption that artificial intelligence and machine learning are "inevitable", such that "'better building' is the only ethical path forward"[71]. In turn, tech ethics efforts pursue technical and procedural solutions for the harmful social consequences of technology[79]. Following this logic, tech companies have developed numerous ethics and fairness toolkits[80–84].

Although efforts to improve the design decisions of individual engineers can be beneficial, the focus on individual design choices relies on a narrow theory of change for how to reform technology. Regardless of their intentions and the design frameworks at their disposal, individual engineers typically have little power to shift corporate strategy. Executives can prevent engineers from understanding the full scope of their work, limiting knowledge and internal dissent about controversial projects[85, 86]. Even when engineers do know about and protest projects, the result is often them resigning or being replaced rather than the company changing course[60, 85]. The most notable improvements in technology use and regulation have come from collective action among activists, tech workers, journalists, and scholars, rather than individual design efforts[87, 88].

More broadly, the emphasis on design ignores the structural sources of technological harms. The injustices associated with digital technologies result from business models that rely on collecting massive amounts of data about the public[89, 90]; companies that wield monopolistic power[91, 92]; technologies that are built through the extraction of natural resources and the abuse of workers[93–96]; and the exclusion of women, minorities, and non-technical experts from technology design and governance[97, 98].

These structural conditions place significant barriers on the extent to which design-oriented tech ethics can guide efforts to achieve reform. As anthropologist Susan Silbey notes, "while we might want to acknowledge human agency and decision-making at the heart of ethical action, we blind ourselves to the structure of those choices—incentives, content, and pattern—if we focus too closely on the individual and ignore the larger pattern of opportunities and motives that channel the actions we call ethics"[78]. To the extent that it defines ethical technology in terms of individual design decisions, tech ethics will divert scrutiny away from the economic and political factors that drive digital injustice, limiting our ability to address these forces.

### 3.3 Tech ethics is subsumed into corporate logics and incentives

Digital technology companies have embraced ethics as a matter of corporate concern, aiming to present the appearance of ethical behavior for scrutinizing audiences. As Alphabet and Microsoft noted in recent SEC filings, products that are deemed unethical could lead to reputational and financial harms[99]. Companies are eager to avoid any backlash, yet do not want to jeopardize their business plans. An ethnography of ethics work in Silicon Valley found that "performing, or even showing off, the seriousness with which a company takes ethics becomes a more important sign of ethical practices than real changes to a product"[79]. For instance, after an effort at Twitter to reduce online harassment stalled, an external researcher involved in the effort noted, "The impression I came away with from this experience is that Twitter was more sensitive to deflecting criticism than in solving the problem of harassment"[100].

Corporate tech ethics is therefore framed in terms of its direct alignment with business strategy. A software engineer at LinkedIn described algorithmic fairness as being profitable for companies, arguing, "If you are very biased, you might only cater to one population, and eventually that limits the growth of your user base, so from a business perspective you actually want to have everyone come on board, so it is actually a good business decision in the long run"[101]. Similarly, one of the people behind the Ethical OS toolkit described being motivated to produce "a tool that helps you think through societal consequences and makes sure what you are designing is good for the world and good for your longer-term bottom line"[102].

Finding this alignment between ethics and business is an important task for those charged with promoting ethics in tech companies. Recognizing that "market success trumps ethics", individuals focused on ethics in Silicon Valley feel pressure to align ethical principles with corporate revenue sources[79]. As one senior researcher in a tech company notes, "the ethics system that you create has to be something that people feel adds value and is not a massive roadblock that adds no value, because if it is a roadblock that has no value, people literally will not do it, because they do not have to"[79]. When ethical ideals are at odds with a company's bottom line, they are met with resistance[1].

This emphasis on business strategy creates significant conflicts with ethics. Corporate business models often rely on extractive and exploitative practices, leading to many of the controversies at the heart of the techlash. Indeed, efforts to improve privacy and curb disinformation have led Facebook and Twitter stock values to decline rapidly[103, 104]. Thus, even as tech companies espouse a devotion to ethics, they continue to develop products and services that raise ethical red flags but promise significant profits. For example, even after releasing AI ethics principles that include safety, privacy, and inclusiveness[31] and committing not to "deploy facial recognition technology in scenarios that we believe will put democratic freedoms at risk"[105], Microsoft invested in AnyVision, an Israeli facial recognition company that supports military surveillance of Palestinians in the West Bank[106]. Similarly, several years after Google withdrew from Project Maven due to ethical concerns among employees, and then created AI ethics guidelines, the company began aggressively pursuing new contracts with the Department of Defense[107].

In sum, tech ethics is being subsumed into existing tech company logics and business practices rather than changing those logics and practices (even if some individuals within companies do want to create meaningful change). This absorption allows companies to take up the mantle of ethics without making substantive changes to their processes or business strategies. The goal in companies is to find practices "which the organization is not yet doing but is capable of doing"[79], indicating an effort to find relatively costless reforms that provide the veneer of ethical behavior. Ethics statements "co-opt the language of some critics", taking critiques grounded in a devotion to equity and social justice and turning them into principles akin to "conventional business ethics"[71]. As they adopt these principles, tech companies "are learning to speak and perform ethics rather than make the structural changes necessary to achieve the social values underpinning the ethical fault lines that exist"[79].

These limits to corporate tech ethics are exemplified by Google's firings of Timnit Gebru and Meg Mitchell. Despite Gebru's and Mitchell's supposed charge as co-leads of Google's Ethical AI team, Google objected to a paper they had written (alongside several internal and external co-authors) about the limitations and harms of large language models, which are central to Google's business[108]. Google attempted to force the authors to retract the paper, claiming that they failed to acknowledge recent technical advances that mitigate many of the paper's concerns[108]. Soon after, journalists revealed that this incident reflected a larger pattern: Google had expanded its review of papers that discuss "sensitive topics", telling researchers, for instance, to "take great care to strike a positive tone" regarding Google's technologies and products[109]. Thus, even as Google publicly advertised its care for ethics, internally the company was carefully reviewing research to curtail ethical criticisms that it deemed threatening to its core business interests.

### 3.4 Tech ethics has become an avenue for ethics-washing

As evidence of tech ethics' limitations has grown, many have critiqued tech ethics as a strategic effort among technology companies to maintain autonomy and profits.

This strategy has been labeled "ethics-washing" (i.e., "ethical white-washing"): adopting the language of ethics to diminish public scrutiny and avoid regulations that would require substantive concessions[110−112]. As an ethnography of ethics in Silicon Valley found, "It is a routine experience at 'ethics' events and workshops in Silicon Valley to hear ethics framed as a form of self-regulation necessary to stave off increased governmental regulation"[79]. This suggests that the previously described issues with tech ethics might be features rather than bugs: by focusing public attention on the actions of individual engineers and on technical dilemmas (such as algorithmic bias), companies perform a sleight-of-hand that shifts structural questions about power and profit out of view. Companies can paint a self-portrait of ethical behavior without meaningfully altering their practices.

Thomas Metzinger, a philosopher who served on the European Commission's High-Level Expert Group on Artificial Intelligence (AI HLEG), provides a particularly striking account of ethics-washing in action[110]. The AI HLEG contained only four ethicists out of 52 total people and was dominated by representatives from industry. Metzinger was tasked with developing "Red Lines" that AI applications should not cross. However, the proposed red lines were ultimately removed by industry representatives eager for a "positive vision" for AI. All told, Metzinger describes the AI HLEG's guidelines as "lukewarm, short-sighted, and deliberately vague" and concludes that the tech industry is "using ethics debates as elegant public decorations for a large-scale investment strategy"[110].

Tech companies have further advanced this "ethics-washing" agenda through funding academic research and conferences. Many of the scholars writing about tech policy and ethics are funded by Google, Microsoft, and other companies, yet often do not disclose this funding[113, 114]. Tech companies also provide funding for prominent academic conferences, including the ACM Conference on Fairness, Accountability, and Transparency (FAccT); the AAAI/ACM Conference on Artificial Intelligence, Ethics, and Society (AIES); and the Privacy Law Scholars Conference (PLSC). Even if these funding practices do not directly influence the research output of individual scholars, they allow tech companies to shape the broader academic and public discourse regarding tech ethics, raising certain voices and conversations at the expense of others.②

In December 2019, then-MIT graduate student Rodrigo Ochigame provided a particularly pointed account of ethics-washing[119]. Describing his experiences working in the Media Lab's AI ethics group and collaborating with the Partnership on AI, Ochigame articulated how "the discourse of 'ethical AI' was aligned strategically with a Silicon Valley effort seeking to avoid legally enforceable restrictions of controversial technologies". Ochigame described witnessing firsthand how the Partnership on AI made recommendations that "aligned consistently with the corporate agenda" by reducing political questions about the criminal justice system to matters of technical consideration. A central part of this effort was tech companies strategically funding researchers and conferences in order to generate a widespread discourse about "ethical" technology. Finding that "the corporate lobby's effort to shape academic research was extremely successful", Ochigame concluded that "big tech money and direction proved incompatible with an honest exploration of ethics".

Ochigame's article prompted heated debate about the value and impacts of tech ethics. Some believed that Ochigame oversimplified the story, failing to acknowledge the many people behind tech ethics[120−122]. On this view, tech ethics is a broad movement that includes efforts by scholars and activists to expose and resist technological harms. Yet many of the people centrally involved in those efforts see their work as distinct from tech ethics. Safiya Noble described Ochigame's article as "All the way correct and worth the time to read"[123]. Lilly Irani and Ruha Benjamin expressed similar sentiments, noting that "AI ethics is not a movement"[124] and that "many of us do not frame our work as 'ethical AI'"[125]. On this view, tech ethics represents the narrow domain of efforts, typically promulgated by tech companies, that explicitly embrace the label of "tech ethics".

The debate over Ochigame's article exposed the fault lines at the heart of tech ethics. The central question is what tech ethics actually entails in practice. While some frame tech ethics as encompassing broad societal debates about the social impacts of technology, others define tech ethics as narrower industry-led efforts to

---

② The integrity of academic tech ethics has been further called into question due to funding from other sources beyond tech companies[115−117]. A related critique of academic tech ethics institutes is the lack of diversity within their leadership[118].

explicitly promote "ethics" in technology. On the former view, tech ethics is an important and beneficial movement for improving digital technology. On the latter view, tech ethics is a distraction that hinders efforts to achieve more equitable technology.

## 4 The Contestation of Tech Ethics

The debates described in the previous section reveal that the central question regarding tech ethics is not whether it is desirable to be ethical, but what "ethics" entails and who gets to define it. Although the label of ethics carries connotations of moral philosophy, in practice the "ethics" in tech ethics tends to take on four overlapping yet often conflicting definitions: moral justice, corporate values, legal risk, and compliance[126]. With all of these meanings conflated in the term ethics, superficially similar calls for tech ethics can imply distinct and even contradictory goals. There is a significant gap between the potential benefits of applying ethics (as in rigorous normative reasoning) to technology and the real-world effects of applying ethics (as in narrow and corporate-driven principles) to technology.

As a result, tech ethics represents a terrain of contestation. The contestation of tech ethics centers on certain actors attempting to claim legitimate authority over what it means for technology to be "ethical", at the expense of other actors. These practices of "boundary-work"[127] enable engineers and companies to maintain intellectual authority and professional autonomy, often in ways that exclude women, minorities, the Global South, and other publics[128–130]. We can see this behavior in technology companies projecting procedural toolkits as solutions to ethical dilemmas, computer scientists reducing normative questions into mathematical metrics, academic tech ethics institutes being funded by billionaires and led primarily by white men, and tech ethics principles being disseminated predominantly by the US and Western Europe. Furthermore, many of the most prominent voices regarding tech ethics are white men who claim expertise while ignoring the work of established fields and scholars, many of whom are women and people of color[131, 132].

Two examples of how ethics has been implemented in other domains—science and business—shed light on the stakes of present debates about tech ethics.

### 4.1 Ethics in science

Many areas of science have embraced ethics in recent decades following public concerns about the social implications of emerging research and applications. Despite the seeming promise of science ethics, however, existing approaches fail to raise debates about the structure of scientific research or to promote democratic governance of science.

Rather than interrogating fundamental questions about the purposes of research or who gets to shape that research, ethics has become increasingly institutionalized, instrumentalized, and professionalized, with an emphasis on filling out forms and checking off boxes[133]. Science ethics bodies suffer from limited "ethical imaginations" and are often primarily concerned with "keeping the wheels of research turning while satisfying publics that ethical standards are being met"[133]. "Ethical analysis that does not advance such instrumental purposes tends to be downgraded as not worthy of public support"[133].

In turn, "systems of ethics play key roles in eliding fundamental social and political issues" related to scientific research[134]. For instance, efforts to introduce ethics into genetic research throughout the 1990s and 2000s treated ethics "as something that could be added onto science—and not something that was unavoidably implicit in it"[134]. The effort to treat ethics as an add-on obscured how "ethical choices inhered in efforts to study human genetic variation, regardless of any explicit effort to practice ethics"[134]. As a result, these research projects "bypassed responsibility for their roles in co-constituting natural and moral orderings of human difference, despite efforts to address ethics at the earliest stages of research design"[134].

The turn to ethics can also entail an explicit effort among scientists to defuse external scrutiny and to develop a regime of self-governance. In the 1970s, frightened by calls for greater public participation in genetic engineering, biologists organized a conference at the Asilomar Conference Center in California[135]. The scientific community at Asilomar pursued two, intertwined goals. First, to present a unified and responsible public image, the Asilomar organizers restricted the agenda to eschew discussions of the most controversial applications of genetic engineering (biological warfare and human genetic engineering).

Second, to convince the American public and politicians that allow biologists could self-govern genetic engineering research, the Asilomar attendees "redefined the genetic engineering problem as a technical one" that only biologists could credibly discuss[135]. Although Asilomar is often hailed as a remarkable occasion of scientific self-sacrifice for the greater good, accounts from the conference itself present a different account. "Self-interest, not altruism, was most evident at Asilomar", as not making any sacrifices and appearing self-serving would have invited stringent, external regulation[135].

Tech ethics mirrors many of these attributes in scientific ethics. As with ethics in other fields of science, tech ethics involves a significant emphasis on institutionalized design practices, often entailing checklists and worksheets. Mirroring ethics in genetic research, the emphasis on ethical design treats ethics as something that can be added on to digital technologies by individual engineers, overlooking the epistemologies and economic structures that shape these technologies and their harms. Just like the molecular biologists at Asilomar, tech companies and computer scientists are defining moral questions as technical challenges in order to retain authority and autonomy.③ The removal of red lines in the European Commission's High-Level Expert Group on AI resembles the exclusion of controversial topics from the agenda at Asilomar.

### 4.2 Corporate ethics and co-optation

Codes of ethics have long been employed by groups of experts (e.g., doctors and lawyers) to codify a profession's expected behavior and to shore up the profession's public reputation[137, 138]. Similarly, companies across a wide range of sectors have embraced ethics codes, typically in response to public perceptions of unethical behavior[139].

Yet it has long been clear that the public benefits of corporate ethics codes are minimal. While ethics codes can help make a group appear ethical, they do little to promote a culture of ethical behavior[139]. The primary goal of business ethics has instead been the "inherently unethical" motivation of corporate self-preservation: to reduce public and regulatory scrutiny by promoting a visible appearance of ethical behavior[139, 140]. Ethics codes promote corporate reputation and profit by making universal moral claims that "are extremely important as claims but extremely vague as rules" and emphasizing individual actors and behaviors, leading to a narrow, "one-case-at-a-time approach to control and discipline"[137]. Ethics codes in the field of information systems have long exhibited a notable lack of explicit moral obligations for computing professionals[74, 141].

Business ethics is indicative of the broader phenomenon of co-optation: an institution incorporating elements of external critiques from groups such as social movements—often gaining the group's support and improving the institution's image—without meaningfully acting on that group's demands or providing that group with decision-making authority[142−144]. The increasing centrality of companies as the target of social movements has led to a particular form of co-optation called "corporatization", in which "corporate interests come to engage with ideas and practices initiated by a social movement and, ultimately, to significantly shape discourses and practices initiated by the movement"[145]. Through this process, large corporate actors in the United States have embraced "diluted and deradicalized" elements of social movements "that could be scaled up and adapted for mass markets"[145]. Two factors make movements particularly susceptible to corporatization: heterogeneity (movement factions that are willing to work with companies gain influence through access to funding) and materiality (structural changes get overlooked in favor of easily commodifiable technological "fixes"). By participating in movement-initiated discourses, companies are able to present themselves as part of the solution rather than part of the problem, and in doing so can avoid more restrictive government regulations.

Tech ethics closely resembles corporate ethics. Abstract and individualized tech ethics codes reproduce the virtue signaling and self-preservation behind traditional business ethics. In a notable example of co-optation and corporatization, technology companies have promoted tech ethics as a diluted and commoditized version of tech-critical discourses that originated among activists, journalists, and critical scholars. Because societal efforts to improve technology are often aimed at companies and include both heterogeneity and materiality, it is particularly vulnerable to

---
③ In an ironic parallel, the Future of Life Institute organized an Asilomar Conference on Beneficial AI in 2017, leading to the development of 23 "Asilomar AI Principles"[136].

corporatization. Through corporatization, tech companies use ethics to present themselves as part of the solution rather than part of the problem and use funding to empower the voices of certain scholars and academic communities. In doing so, tech companies shore up their reputation and hinder external regulation. The success of tech ethics corporatization can be seen in the expanding scope of work that is published and discussed under the banner of "tech ethics". Even scholars who do not embrace the tech ethics label are increasingly subsumed into this category, either lumped into it by others or compelled into it as opportunities to publish research, impact policymakers, and receive grants are increasingly shifting to the terrain of "tech ethics".

### 4.3 The stakes of tech ethics

These examples of ethics in science and business suggest two conclusions about tech ethics. First, tech ethics discourse enables technologists and technology companies to label themselves as "ethical" without substantively altering their practices. Tech ethics follows the model of science ethics and business ethics, which present case studies for how ethics-washing can stymie democratic debate and oversight. Continuing the process already underway, tech companies and technologists are poised to define themselves as "ethical" even while continuing to generate significant social harm. Although some individuals and groups are pursuing expansive forms of tech ethics, tech companies have sufficient influence to promote their narrow vision of "tech ethics" as the dominant understanding and implementation.

Second, those striving for substantive and structural improvements in digital technologies must be mindful of the gap between ethics as normative inquiry and ethics as a practical endeavor. Moral philosophy is essential to studying and improving technology, suggesting that ethics is inherently desirable. However, the examples of ethics in technology, science, and business indicate that ethics in practical contexts can be quite distinct from ethics as a mode of moral reasoning. It is necessary to recognize these simultaneous and conflicting roles of ethics. Defenders of ethics-as-moral-philosophy must be mindful not to inadvertently legitimize ethics-as-superficial-practice when asserting the importance of ethics. Meanwhile, critics who would cede ethics to tech companies and engineers as a denuded concept should be mindful that ethics-as-moral-philosophy has much to offer their own critiques of ethics-as-superficial-practice.

Attending to these porous and slippery boundaries is essential for supporting efforts to resist oppressive digital technologies. As indicated by the responses to Ochigame's critique of ethics-washing, many of the more radical critics of digital technology see themselves as outside of—if not in opposition to—the dominant strains of tech ethics. Activists, communities, and scholars have developed alternative discourses and practices: refusal[85, 146, 147], resistance[148], defense[149, 150], abolition[150, 151], and decentering technology[152]. Although some may see these alternative movements as falling under the broad umbrella of tech ethics, they embody distinct aspirations from the narrow mainstream of tech ethics. Labeling these burgeoning practices as part of tech ethics risks giving tech ethics the imprimatur of radical, justice-oriented work even as its core tenets and practices eschew such commitments.

## 5 A Sociotechnical Approach to Tech Ethics

Rather than presenting a unifying and beneficent set of principles and practices, tech ethics has emerged as a central site of struggle regarding the future of digital architectures, governance, and economies. Given these dynamics of contestation surrounding tech ethics, ethics will not, on its own, provide a salve for technology's social harms. In order to better evaluate the opportunities and limits of tech ethics, it is necessary to shift our focus from the value of ethics in theory to the impacts of ethics in practice.

This task calls for analyzing tech ethics through a sociotechnical lens. A sociotechnical approach to technology emphasizes that artifacts cannot be analyzed in isolation. Instead, it is necessary to focus on technology's social impacts and on how artifacts shape and are shaped by society. Similarly, a sociotechnical approach to tech ethics emphasizes that tech ethics cannot be analyzed in isolation. Instead, it is necessary to focus on the social impacts of tech ethics and on how tech ethics shapes and is shaped by society. If "technologies can be assessed only in their relations to the sites of their production and use"[22], then so too, we might say, tech ethics can be assessed only in relation to the sites of its conception and use. With this aim in mind, it is fruitful to consider tech ethics through the lens of four sociotechnical frames: objectivity and neutrality, determinism, solutionism, and sociotechnical systems.

## 5.1 Objectivity and neutrality

A sociotechnical lens on technology sheds light on how scientists and engineers are not objective and on how technologies are not neutral. It makes clear that improving digital technologies requires grappling with the normative commitments of engineers and incorporating more voices into the design of technology[153, 154]. Similarly, it is necessary to recognize that the actors promoting tech ethics are not objective and that tech ethics is not neutral. Currently, the range of perspectives reflected in ethics principles is quite narrow and ethics is treated as an objective, universal body of principles[29, 71, 73]. In many cases, white and male former technology company employees are cast to the front lines of public influence regarding tech ethics[131, 132]. As a result, the seeming consensus around particular ethical principles may say less about the objective universality of these ideals than about the narrow range of voices that influence tech ethics. Thus, rather than treating tech ethics as a body of objective and universal moral principles, it is necessary to grapple with the standpoints and power of different actors, the normative principles embodied in different ethical frameworks, and potential mechanisms for adjudicating between conflicting ethical commitments.

## 5.2 Determinism

A central component of a sociotechnical approach to technology is rejecting technological determinism: the belief that technology evolves autonomously and determines social outcomes[155, 156]. Scholarship demonstrates that even as technology plays a role in shaping society, technology and its social impacts are also simultaneously shaped by society[21, 23, 157, 158]. Similarly, it is necessary to recognize the various factors that influence the impacts of tech ethics in practice. Currently, ethics in digital technology is often treated through a view of "ethical determinism", with an underlying assumption that adopting "ethics" will lead to ethical technologies. Yet evidence from science, business, and digital technology demonstrates that embracing "ethics" is typically not sufficient to prompt substantive changes. As with technology, ethics does not on its own determine sociotechnical outcomes. We therefore need to consider the indeterminacy of tech ethics: i.e., how the impacts of tech ethics are shaped by social, political, and economic forces.

## 5.3 Solutionism

Closely intertwined with a belief in technological determinism is the practice of technological solutionism: the expectation that technology can solve all social problems[159]. A great deal of sociotechnical scholarship has demonstrated how digital technology "solutions" to social problems not only typically fail to provide the intended solutions, but also can exacerbate the problems they are intended to solve[160−163]. Similarly, it is necessary to recognize the limits of what tech ethics can accomplish. Currently, even as tech ethics debates have highlighted how technology is not always the answer to social problems, a common response has been to embrace an "ethical solutionism": promoting ethics principles and practices as the solution to these sociotechnical problems. A notable example (at the heart of many tech ethics agendas) is the response to algorithmic discrimination through algorithmic fairness, which often centers narrow mathematical definitions of fairness but leaves in place the structural and systemic conditions that generate a great deal of algorithmic harms[164, 165]. Efforts to introduce ethics in digital technology function similarly, providing an addendum of ethical language and practices on top of existing structures and epistemologies which themselves are largely uninterrogated. Thus, just as technical specifications of algorithmic fairness are insufficient to guarantee fair algorithms, tech ethics principles are insufficient to guarantee ethical technologies. Ethics principles, toolkits, and training must be integrated into broader approaches for improving digital technology that include activism, policy reforms, and new engineering practices.

## 5.4 Sociotechnical systems

A key benefit of analyzing technologies through a sociotechnical lens is expanding the frame of analysis beyond the technical artifact itself. Rather than operating in isolation, artifacts are embedded within sociotechnical systems, such that the artifact and society "co-produce" social outcomes[21]. Similarly, it is necessary to view tech ethics as embedded within social, economic, and legal environments, which shape the uses and impacts of tech ethics. Currently, efforts to promote ethical technology typically focus on the internal characteristics of tech ethics—which principles to promote, for instance—with little attention to the impacts of these efforts when integrated into a tech company or computer science curriculum. In turn, tech ethics has had limited effects on technology production and has played a legitimizing role for technology companies. Attempts to promote more equitable technology must instead consider the full context in which tech ethics is embedded. The impacts of tech ethics are shaped by the beliefs and actions of engineers, the economic incentives of companies, cultural and political pressures, and regulatory environments. Evaluating tech ethics in light of these factors can generate better predictions about how particular efforts will fare in practice. Furthermore, focusing on these contextual factors can illuminate reforms that are more likely to avoid the pitfalls associated with tech ethics.

## 6 Conclusion

A sociotechnical lens on tech ethics will not provide clear answers for how to improve digital technologies. The technological, social, legal, economic, and political challenges are far too entangled and entrenched for simple solutions or prescriptions. Nonetheless, a sociotechnical approach can help us reason about the benefits and limits of tech ethics in practice. Doing so will inform efforts to develop rigorous strategies for reforming digital technologies.

That is the task of this special issue: "Technology Ethics in Action: Critical and Interdisciplinary Perspectives". The articles in this issue provide a range of perspectives regarding the value of tech ethics and the desirable paths forward. By interrogating the relationships between ethics, technology, and society, we hope to prompt reflection, debate, and action in the service of a more just society.

## Acknowledgment

B. Green thanks Elettra Bietti, Anna Lauren Hoffmann, Jenny Korn, Kathy Pham, and Luke Stark for their comments on this article. B. Green also thanks the Harvard STS community, particularly Sam Weiss Evans, for feedback on an earlier iteration of this article.

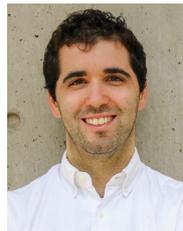

**Ben Green** is a postdoctoral scholar in the Society of Fellows and an assistant professor in the Gerald R. Ford School of Public Policy, University of Michigan. He received the PhD degree in applied math (with a secondary field in STS) from Harvard University and the BS degree in mathematics & physics from Yale College in 2020 and 2014, respectively.